\documentclass[10pt,twocolumn,letterpaper]{article}
\usepackage{spconf,amsmath,graphicx}
\usepackage{amsmath}
\usepackage{amssymb}
\usepackage{float}
\usepackage{graphicx}
\usepackage[dvipsnames]{xcolor}
\usepackage{mathtools}
\usepackage{subfigure}
\usepackage{enumitem}

\DeclarePairedDelimiter\abs{\lvert}{\rvert}%
\DeclarePairedDelimiter\norm{\lVert}{\rVert}%

\makeatletter
\let\oldabs\abs
\def\abs{\@ifstar{\oldabs}{\oldabs*}}
\let\oldnorm\norm
\def\norm{\@ifstar{\oldnorm}{\oldnorm*}}

\makeatother
\usepackage{amsmath}
\usepackage{siunitx}
\makeatletter
\providecommand\add@text{}
\newcommand\tagaddtext[1]{%
  \gdef\add@text{#1\gdef\add@text{}}}%
\renewcommand\tagform@[1]{%
  \maketag@@@{\llap{\add@text\quad}(\ignorespaces#1\unskip\@@italiccorr)}%
}
\makeatother

\def\@fnsymbol#1{\ensuremath{\ifcase#1\or *\or \dagger\or \ddagger\or
   \mathsection\or \mathparagraph\or \|\or **\or \dagger\dagger
   \or \ddagger\ddagger \else\@ctrerr\fi}}
\newcommand{\ssymbol}[1]{^{\@fnsymbol{#1}}}

\title{On Psychoacoustically Weighted Cost Functions Towards Resource-Efficient Deep Neural Networks for Speech Denoising}
\name{Kai Zhen$^1$, Aswin Sivaraman$^1$, Jongmo Sung$^2$, Minje Kim$^1$\thanks{This work was supported by Institute for Information \& communications Technology Promotion(IITP) grant funded by the Korea government (MSIT) (2017-0-00072, Development of Audio/Video Coding and Light Field Media Fundamental Technologies for Ultra Realistic Tera-media).}}
\address{$^1$Indiana University, School of Informatics, Computing, and Engineering, Bloomington, IN\\
$^2$Electronics and Telecommunications Research Institute, Daejeon, South Korea\\
  {\small \tt zhenk@umail.iu.edu, asivara@iu.edu, minje@indiana.edu, jmseong@etri.re.kr}}

\begin{document}
\ninept
\maketitle
\begin{abstract}
We present a psychoacoustically enhanced cost function to balance network complexity and perceptual performance of deep neural networks for speech denoising. While training the network, we utilize perceptual weights added to the ordinary mean-squared error to emphasize contribution from frequency bins which are most audible while ignoring error from inaudible bins. To generate the weights, we employ psychoacoustic models to compute the global masking threshold from the clean speech spectra. We then evaluate the speech denoising performance of our perceptually guided neural network by using both objective and perceptual sound quality metrics, testing on various network structures ranging from shallow and narrow ones to deep and wide ones. The experimental results showcase our method as a valid approach for infusing perceptual significance to deep neural network operations. In particular, the more perceptually sensible enhancement in performance seen by simple neural network topologies proves that the proposed method can lead to resource-efficient speech denoising implementations in small devices without degrading the perceived signal fidelity.
\end{abstract}


\section{Introduction}
\label{sec:intro}

Deep Neural Networks (DNNs) have seen exponentially greater usage with regards to audio signal processing, improving the state-of-the-art in source separation, noise reduction, and speech enhancement. In many of these studies, their improved performance in terms of the quality recovery of the sources relies greatly on the enlarged model complexity. For example, a network with $300\times2$ structure (2 hidden layers, each of which has 300 units) showed speech separation performance more than 1 dB better than a traditional dictionary-based separation model (where the dictionaries are learned from Non-negative Matrix Factorization (NMF) \cite{LeeDD99nature, LeeDD2000nips} in advance) in terms of Signal-to-Distortion Ratio (SDR) \cite{HuangP2015ieeeacmaslp}. Another recent example would be a DNN with $1024\times3$ structure where both phase and magnitudes of the source are effectively predicted \cite{WilliamsonD2017jasa}. If the network was represented as a weight matrix in a linear transformation, the required number of floating-point operations would easily be over a few million. In \cite{LeRouxJ2015icassp} it is also shown that for standard feed-forward networks, a larger network structure ($1024\times2$ with 3.1M parameters) outperforms a smaller network ($256\times3$ with 644K parameters) by 1.17 dB. As expected, the network complexity is predominantly increased in favor of improved performance. However, the enlarged structure can become a bottleneck when it comes to implementing the DNN in a small device with limited resources (e.g. power and memory), especially when there is a stringent requirement for real-time speech enhancement.


As DNNs increase in their size and resource usage, neural network compression has grown to be a lively research area. Carefully pruning some of the units can reduce the size of the network as shown in \cite{HanS2016iclr}. Lowering the quantization level for the network parameters by reducing the number of bits to represent each parameter is another way to compress the network. For example, recent studies report that binary or ternary quantization schemes do not significantly reduce accuracy in famous benchmark classification tasks \cite{HwangK2014sips, KimMJ2015icmlw, HubaraI2016nips, RastegariMCoRR16}. However, those general-purpose compression techniques do not utilize the audio-specific characteristics of the problem. For example, as proposed in the Perceptual Evaluation methods for Audio Source Separation (PEASS) toolkit \cite{peass}, standard energy-based objective metrics such as Signal-to-Noise Ratio (SNR) and its variants are not the best way to judge the perceptual quality of audio signals \cite{VincentE2006ieeeaslp}).

In this paper we claim that a neural network for audio enhancement can be optimized further if it exploits human perception. For example, legacy digital audio coding schemes leverage principles of sound perception to reduce the coded signal's data rate; we refer to these perceptual coding schemes as Psychoacoustic Models (PAM). By discarding inaudible tones and allowing more quantization noise in the least audible portions of the audio spectrum, PAM reduces the bit rate while minimizing degradation of the overall signal fidelity \cite{bosi}. The psychoacoustic concepts most relevant to the design of a perceptual audio coder are the phenomena of hearing thresholds and of auditory masking. These concepts have been empirically modeled and can be used in conjunction with time-frequency analysis to identify the perceptually significant (i.e. tonal and non-tonal) components of the signal. Once identified, auditory irrelevancies, which are either masked or unheard, can be removed. Because psychoacoustics literature is diverse in its findings, different modern perceptual audio codecs have adopted their own PAM. We will incorporate PAM-1 \cite{painter}, popularized by the ISO/IEC MPEG-1 standard without the loss of generality. 

Although PAM has been commonly used for traditional audio coding, it has not yet been used with respect to neural network compression. Recent work proposed by Liu et al. regarding a perceptually weighted DNN for speech enhancement shows a promising amount of improvement in both the signal quality and intelligibility \cite{liu}. However, the network uses a perceptual weight model based on a sigmoid function applied to the signal, which prioritizes high-energy components of the target speech's spectral power. Although this assumption is valid in the case of speech, unlike PAM it does not hold across all types of audio. In this regard, our work generalizes and greatly expands upon the findings found by Liu et al. 

In this paper, we present a perceptually weighted cost function to train a DNN that is structurally simpler, but conducts perceptually comparable speech denoising. To do this, we will generate meaningful weights based on the global masking threshold of our training data as prescribed by PAM-1, and then harmonize the weights with the mean-squared error. We evaluate denoising results from various network architectures and show that the proposed method leads to a more condensed network topology without losing the perceptual quality of the recovered speech.

\section{Background}
\label{sec:model}


\subsection{Conventional Mask Learning Networks}
The input of the mask learning model is the magnitude spectra of the noisy utterances $\mathbf{X}$ that approximates the mixture of speech and noise in the complex domain: $\mathbf{X}\approx\mathbf{S}+\mathbf{N}$\footnote{We assume all data matrices are magnitude spectrograms, however the exact mixture is defined in the complex time-frequency domain}. Rectified Linear Units (ReLU) are common as the activation function to avoid the gradient vanishing problem. For the $(i+1)$-th hidden layer, the feed-forward process is defined as follow: 
\begin{equation}
\label{eq:alayer}
\mathbf{X}^{(i+1)} = \sigma_{\text{ReLU}}\!\left(\tanh\!\big(\mathbf{W}^{(i)}\big)\!\left(\mathbf{R}^{(i)}\odot\! \mathbf{X}^{(i)}\!\right)\!+\!\tanh\big(\mathbf{b}^{(i)}\big)\!\right)\!,
\end{equation}
where $i$ indicates the layers with $i=0$ as the special case for the input layer (i.e. $\mathbf{X}^{(0)}$ stands for the input), and $\mathbf{W}^{(i)}$ and $\mathbf{b}^{(i)}$ are for the weights and bias, respectively. $\odot$ denotes Hadamard product. In practice, we find it useful to conduct a smooth weight clipping by applying the hyperbolic tangent function to each weight and bias \eqref{eq:alayer}, which will be bounded within the range of $-1$ to $+1$. Dropout is applied to the output of a layer before it is fed to the next layer; for dropout, we use $\mathbf{R}^{(i)}$ as the masking matrix for the $i$-th layer whose elements are binary values drawn from a Bernoulli distribution with parameter $\rho^{(i)}$.  


The speech denoising networks are trained to predict the Ideal Ratio Mask (IRM) $\mathbf{M}$, which can mask the input to recover the speech spectrogram $\mathbf{S}=\mathbf{M}\odot\mathbf{X}$ \cite{NarayananA2013icassp}. Hence, we employ the logistic function as the final layer activation, modifying \eqref{eq:alayer} as follows:
\begin{equation}\nonumber
\mathbf{X}^{(L+1)}=\sigma\Big( \tanh\big(\mathbf{W}^{(L)}\big)\left(\mathbf{R}^{(L)}\odot \mathbf{X}^{(L)}\right)+\tanh\big(\mathbf{b}^{(L)}\big) \Big)
\end{equation}
From this we define the mean-squared error function between the IRM and the network output:
\begin{equation}
\label{loss1}
\mathcal{E}\left(\mathbf{M}||\mathbf{X}^{(L+1)}\right)= \frac{1}{FT}\sum_{t=1}^{T}\sum_{f=1}^{T}(\mathbf{X}^{(L+1)}_{f,t}-\mathbf{M}_{f,t})^2
\end{equation}
where $F$ and $T$ are the number of input dimensions and the number of samples in the training data, respectively.

%

%
%
%
%
%




%

\begin{figure}[t]
\centering
    \includegraphics[width=\columnwidth]{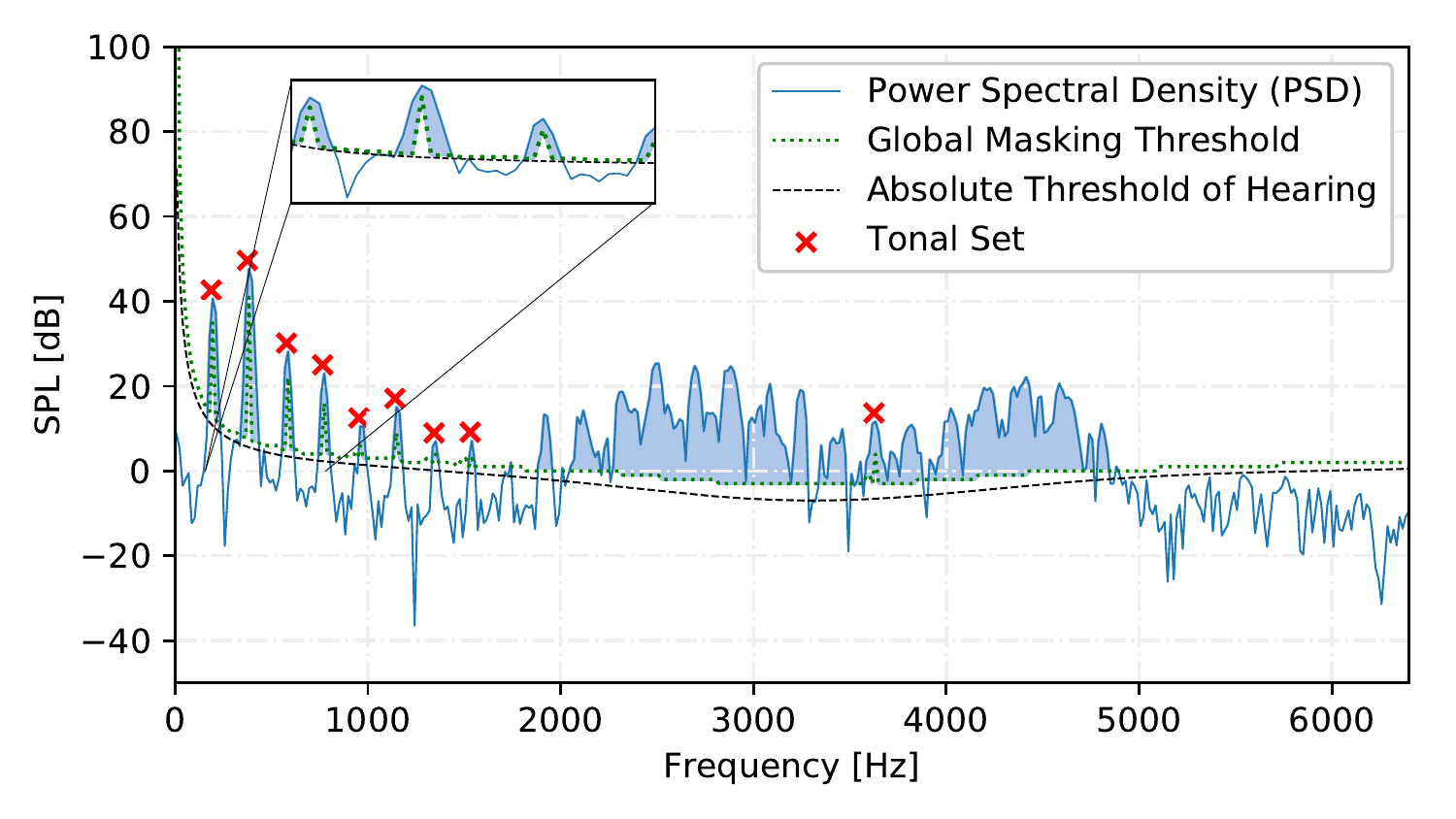}
 \caption{Plot of pyschoacoustic model components. PAM-1 identifies the tonal set ( red crosses) and then estimates a global masking threshold (green dotted line) on top of the absolute treshold of hearing (black dashed line). These components are all determined using the input signal's power spectral density (blue line). The shaded area between the PSD curve and the global masking threshold represents audible spectral energy.} 
 \label{fig:psycho-model}
\end{figure}

\subsection{Psychoacoustic Models}

Incorporation of a psychoacoustic model is essential to constructing a weight matrix which influences the cost function to focus on signal components of greatest perceptual significance. We utilize a simplified version of PAM-1, by only considering the tonal signal components and ignoring the non-tonal (i.e. noise-making) signal components. PAM-1 computes the global masking threshold for all frames of the input spectrogram. The threshold is computed first by performing a Sound Pressure Level (SPL) normalization of the training spectrogram ($\mathbf{S}$) to determine the signal's Power Spectral Density (PSD): $\mathbf{P}=P_N + 10 \log_{10}\abs*{\mathbf{S}}^2$. Note that $P_N$ is fixed to $90.302$ dB. Then, the model identifies the tonal maskers, ignoring those which fall below the absolute threshold of hearing (ATH). A spreading function is used to generate masking curves for each tone. The combination of these individual masking curves plus the ATH yields the global masking threshold $\mathbf{G}$. This implementation of PAM-1 is detailed in \cite{painter}. Fig. \ref{fig:psycho-model} showcases the various components of the PAM-1 model for an example frame from the training data set.

\section{Proposed Perceptual Weighting}

The proposed method reformulate a given ordinary cost function by using the perceptual weights derived from the masking curves computed from PAM-1. Using the training clean speech signal's power spectral density ($\mathbf{P}$) and its corresponding global masking threshold ($\mathbf{G}$), we define a perceptual weight matrix ($\mathbf{H}$) which is applied to the network cost function \eqref{loss1}:
\begin{equation}
    \mathbf{H}=\log_{10}\left(\frac{10^{0.1\mathbf{P}}}{10^{0.1\mathbf{G}}}+1\right),
\end{equation}
Therefore, $\mathbf{H}$ is the log ratio between the signal power and the masked threshold rescaled from dB-SPL. Division is carried out in the element-wise fashion. The intuition behind this weight matrix definition can be understood by observing Fig. \ref{fig:psycho-model}. For any signal energy in frequency bin $f$ of the $t$-th time frame, if the signal's power is greater than its masking threshold, i.e. $\mathbf{P}_{f,t}>\mathbf{G}_{f,t}$, this tone must be audible. In Fig. \ref{fig:psycho-model}, the audible regions are those where the blue line is higher than the green dotted line. On the other hand, if the power of the source spectrum is lower than the threshold, the region is masked and inaudible. With this understanding, we define weights bounded between 0 and $\infty$, whose smaller extreme says that the masking threshold is very large and any sound, such as the reconstruction error at that time-frequency bin, is not audible. Conversely, a large weight value means that the source spectral component is large and audible even considering the masking threshold. Therefore, the system should not create much error. Now that the weight matrix $\mathbf{H}$ encodes the perceptual importance of all time-frequency bins, we combine it with the original MSE function:
\begin{equation}
\mathcal{E}\left(\mathbf{M}||\mathbf{X}^{(L+1)}\right)= \frac{1}{FT}\sum_{t=1}^{T}\sum_{f=1}^{T}\mathbf{H}_{f,t}(\mathbf{X}^{(L+1)}_{f,t}-\mathbf{M}_{f,t})^2.
\label{eq:error-pam}
\end{equation}

\begin{figure}[t]
\centering
    \subfigure[]{\includegraphics[height=1.9in]{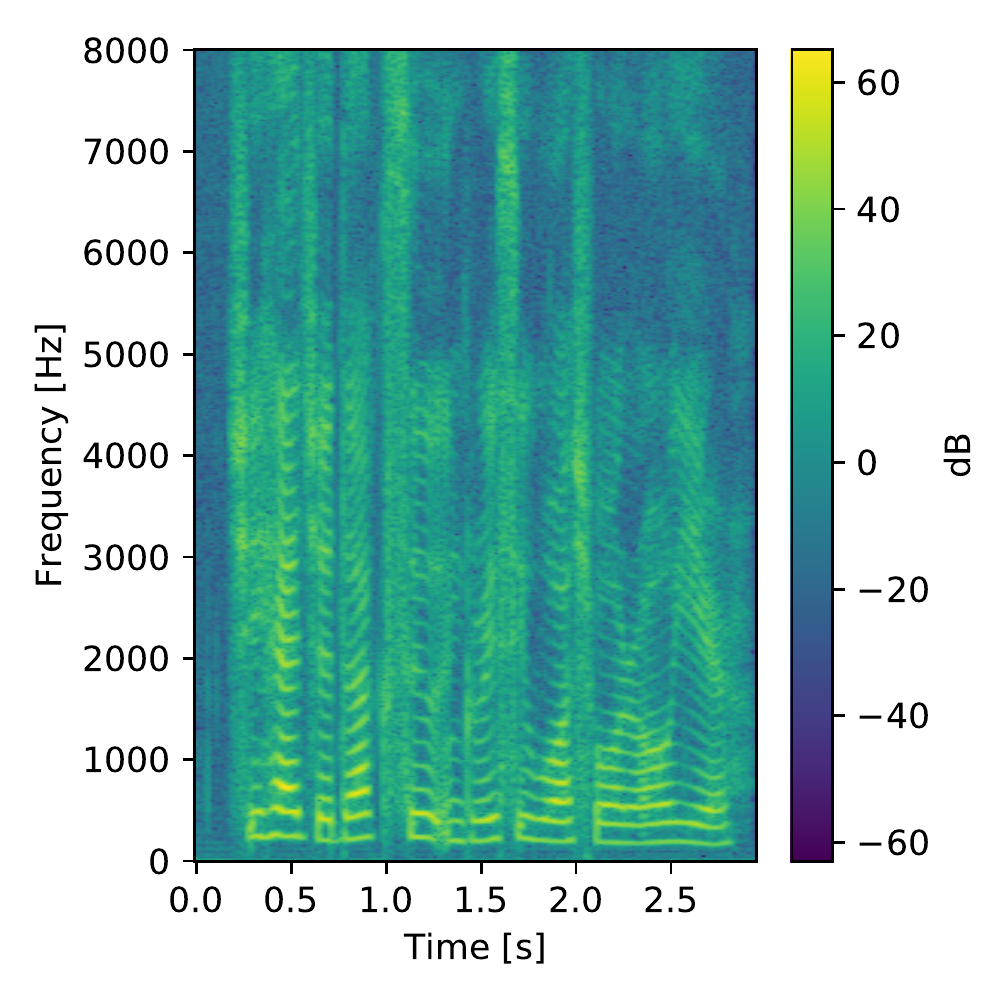}}~~
    \subfigure[]{\includegraphics[height=1.9in]{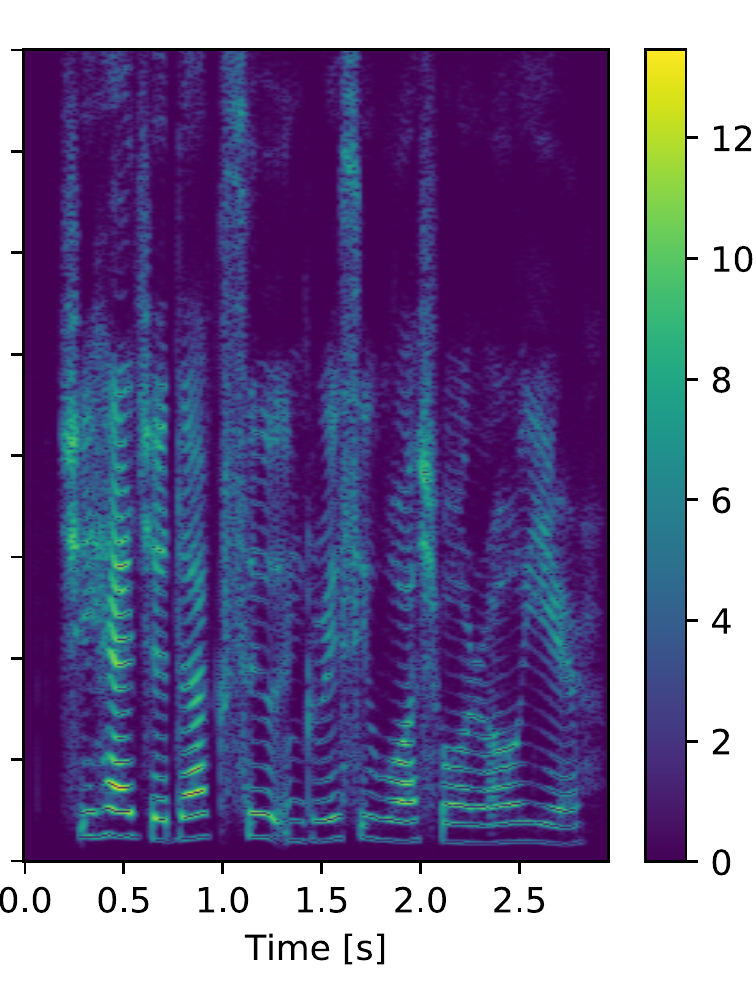}}
    \caption{Comparison of power spectral density ($\mathbf{P}$) (a) against perceptual weight matrix ($\mathbf{H}$) (b) for a single training set speech utterance.}
    \label{fig:psycho-weights}
\end{figure}
Figure \ref{fig:psycho-weights} shows the training signal's power spectral density $\mathbf{P}$ compared to the perceptual weight matrix $\mathbf{H}$ for a short speech signal. Note that $\mathbf{H}$ roughly follows the spectral density while suppressing weaker areas with near-zero weights.

The expected benefit of using this weighting scheme is that the neural network prediction can enjoy a relaxed version of the error function that underweights the less audible output dimensions. As a result, the network can focus more on the narrower output dimensions--an easier optimization task that a smaller and compressed network can also solve with similar perceptual quality.

\section{Experimental Setup}

\begin{itemize}[leftmargin=0in]
\setlength{\itemindent}{.15in}
\item \textbf{\em Data Preparation}: The noisy dataset is constructed by mixing utterances from TIMIT corpus and ten non-stationary noise types used in \cite{noise}. For this, $10$ speakers are randomly selected from the training set with equal gender probability. Each utterance is mixed with one of ten different noise types. Thus, $1,000$ noisy utterances are used for training. This same procedure is used to generate $200$ noisy utterances for validation as well as another $400$ utterances to create the test set. Noise signals used for training do not overlap with those for test mixtures.


Sources are normalized so that the mixture SNR is $0$dB. The Short-Time Fourier Transform (STFT) is used with a $1024$-point Hann window and a $75\%$-overlap for all spectrogram computation. The complex spectrogram of the clean signal ($\mathbf{S}$) and the background noise ($\mathbf{N}$) are mixed to create matrices of dimensions $513\times199261$ for the training set and $513\times34385$ for the validation set. The input mixture to be denoised is acquired by adding up $\mathbf{S}$ and $\mathbf{N}$. For the larger networks (with 1024 or 2048 hidden units per layer) three consecutive spectra are then concatenated and vectorized to form an input vector of $513\times3$. Vectorizing consecutive input is common practice to provide contextual information to fully connected neural networks. For the smaller networks (with 128 or 512 hidden units per layer), the individual frames are the input. The mini-batch size is 256 throughout the training procedure.

The energy-based Ideal Ratio Mask (IRM) gives us a nonnegative real-valued masking matrix $\mathbf{M} = \frac{|\mathbf{S}|^2}{|\mathbf{S}|^2+|\mathbf{N}|^2}$. The source spectrogram is recovered by multiplying the mask to the mixture spectrogram, i.e. $\mathbf{S} \approx \mathbf{M} \odot \mathbf{X}$. We chose IRM to be our target signal, but the proposed perceptual weighting can be used for other targets, such the source magnitude spectra, without the loss of generality. 

\item \textbf{\em Parameter Settings}: 
As we seek a condensed network structure, we limit the maximum number of hidden layers to be $3$, each of which can have $128$, $512$, $1024$, and $2048$ units, totalling $12$ different network topologies assessed in this experiment. The weights per layer are initialized from the truncated normal distribution divided by the square root of the size of the layer, where the standard deviation is set to be $0.1$. We use MSE as our cost function. Learning rates are set to be $10^{-5}$ or $10^{-6}$ given different model topologies. For the dropout rate on the input layer, we choose $\rho^{(0)}=0.95$ to keep most of the units active in most models, but for the concatenated input layer we use $0.5$. Dropout rate is set to be $\rho^{(l)}=0.6$ for the ``wider" hidden layers ($2048$ units), $0.7$ for layers with $1024$ units and $0.8$ for $512$ units. Due to a limited space, the paper only shows the results from the best-fit parameter configuration which we found via validation. Each choice of model structure is trained with two different error functions: with and without the perceptual weights.


\item \textbf{\em Model Training}: 
The model is trained in TensorFlow using the Adam optimizer \cite{tensorflow}. The number of epochs is set to be $5,000$. 
The training time varies in a range between $14$ to $28$ hours given different network structures and parameter settings, but the proposed perceptual weighting does not notably reduce training time. 


\end{itemize}
\label{sec:data}

\begin{figure*}[t]
\centering{
    \subfigure[SDR]{\includegraphics[height=1.8in,trim={.7cm 0 .7cm 0}]{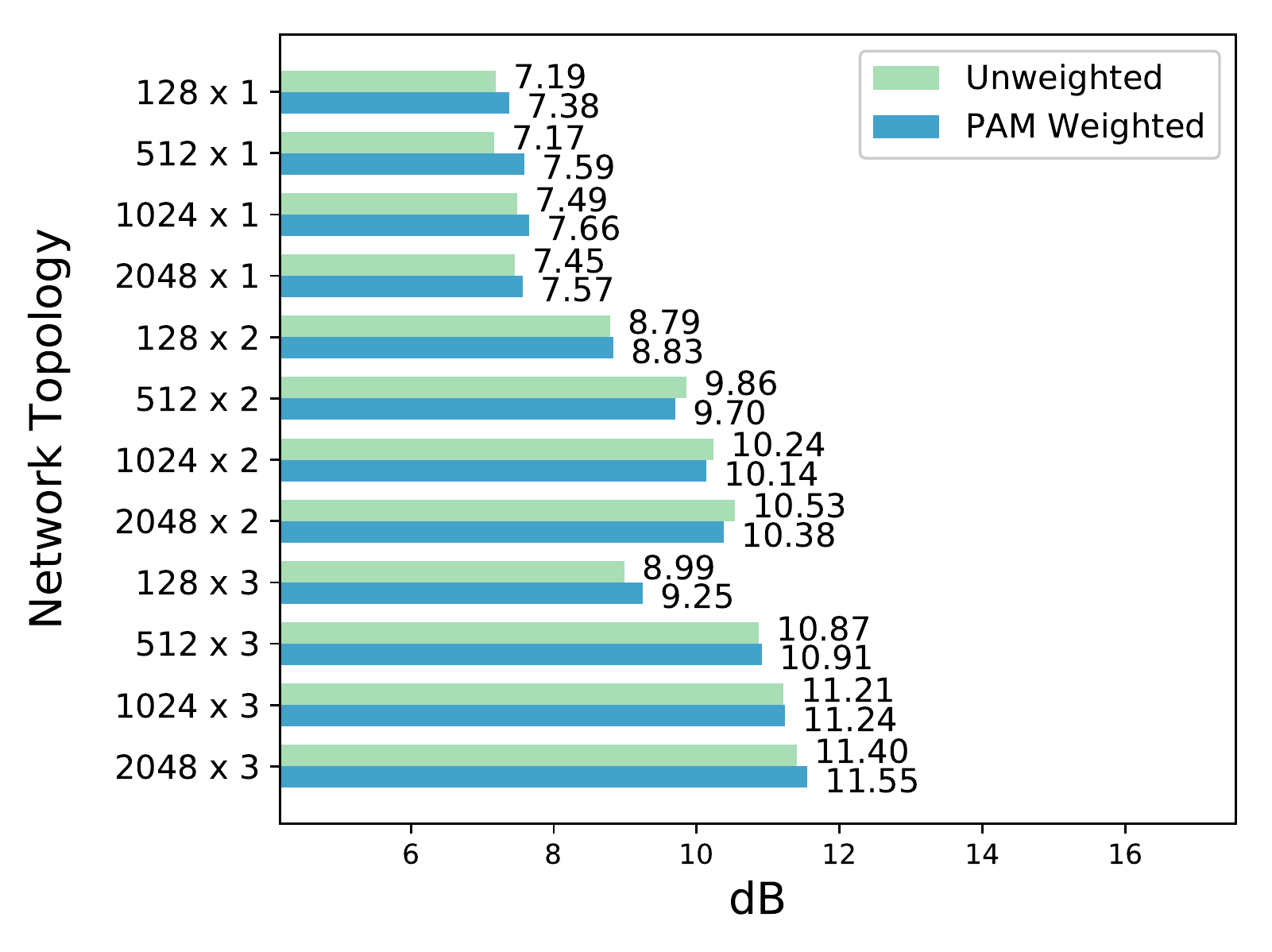}}
    \subfigure[SIR]{\includegraphics[height=1.8in,trim={0 0 .7cm 0}]{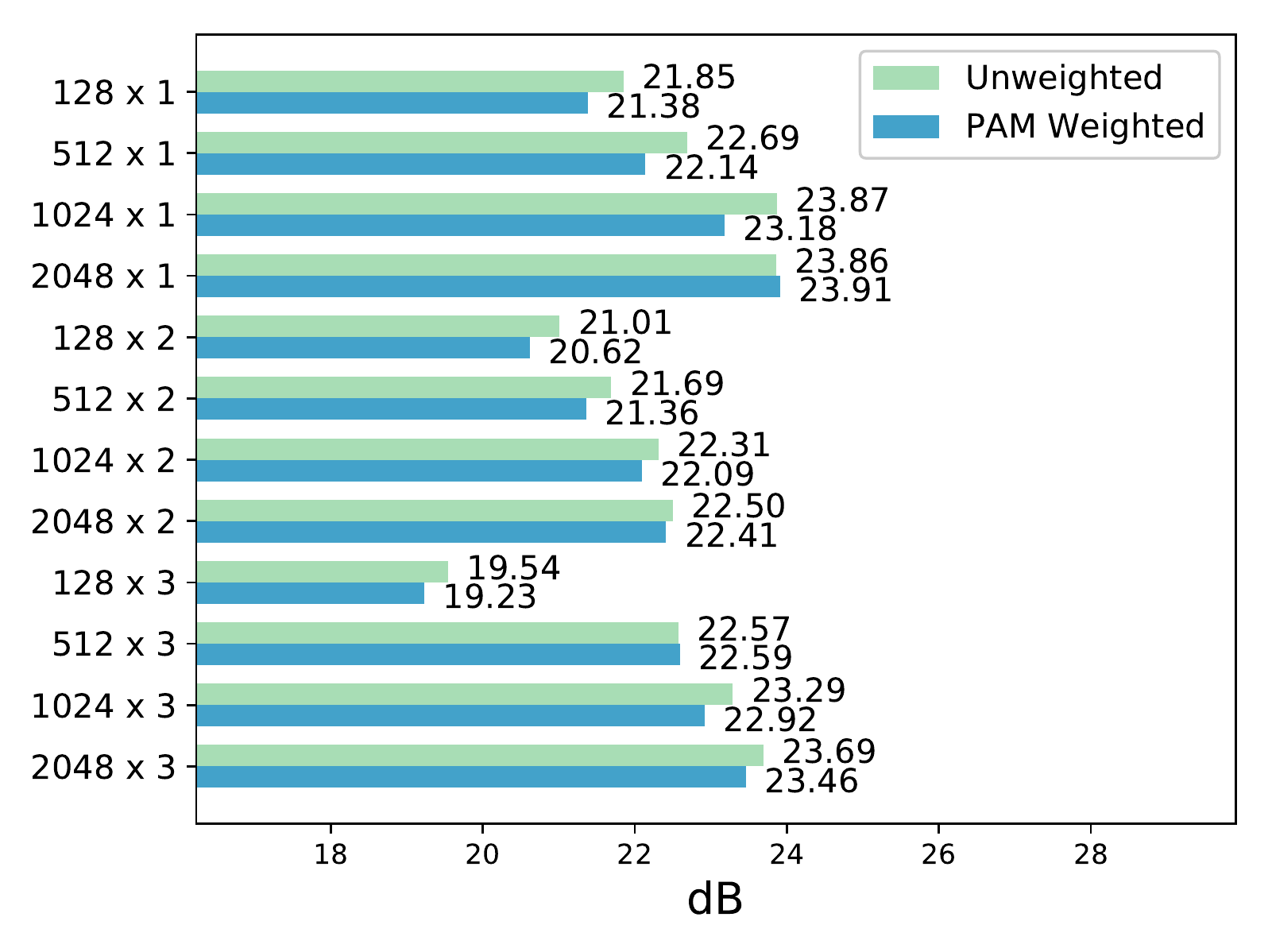}}
    \subfigure[SAR]{\includegraphics[height=1.8in,trim={0 0 .7cm 0}]{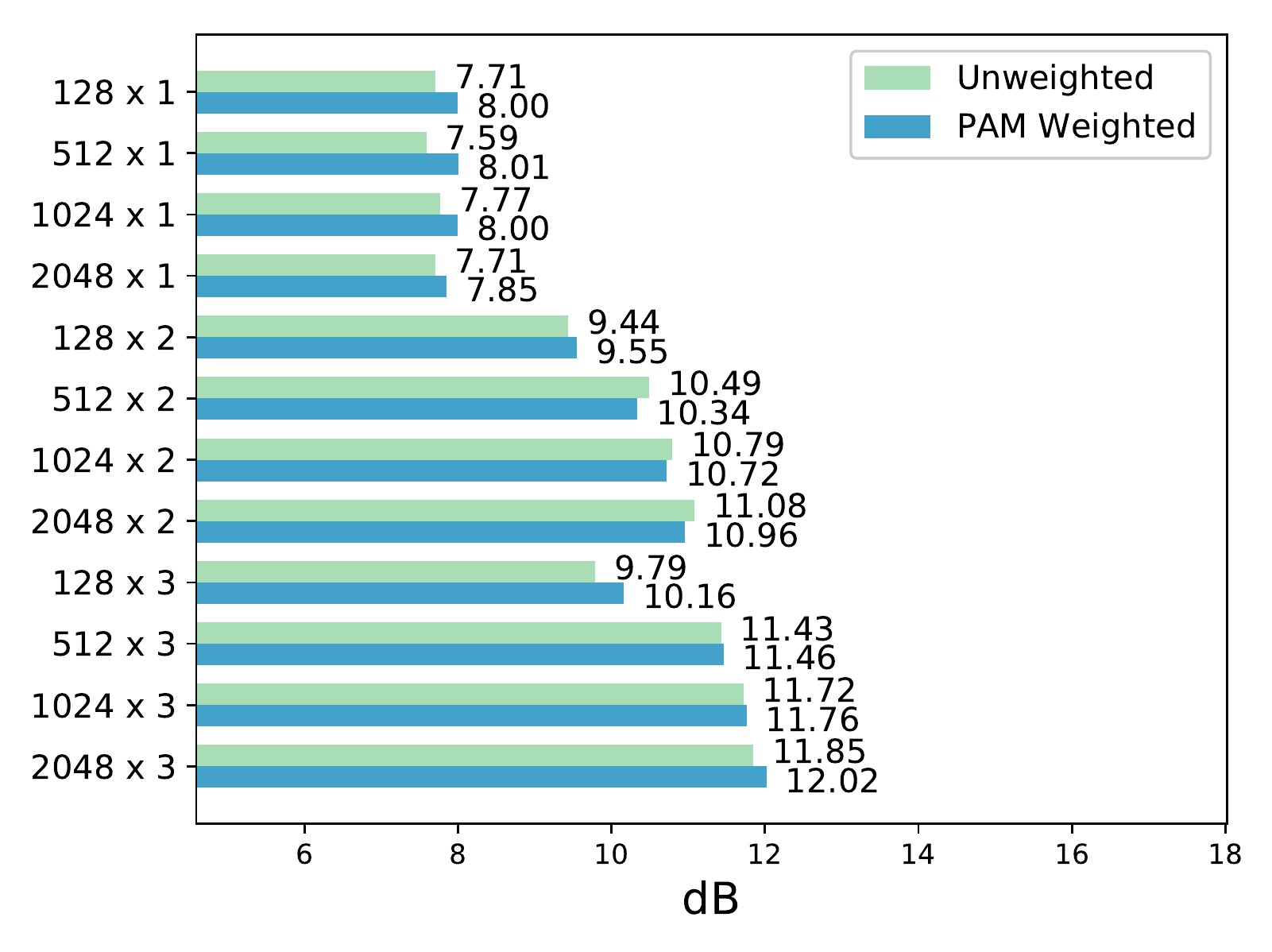}}}
\centering{    
    \subfigure[OPS]{\includegraphics[height=1.8in,trim={.7cm 0 .7cm 0}]{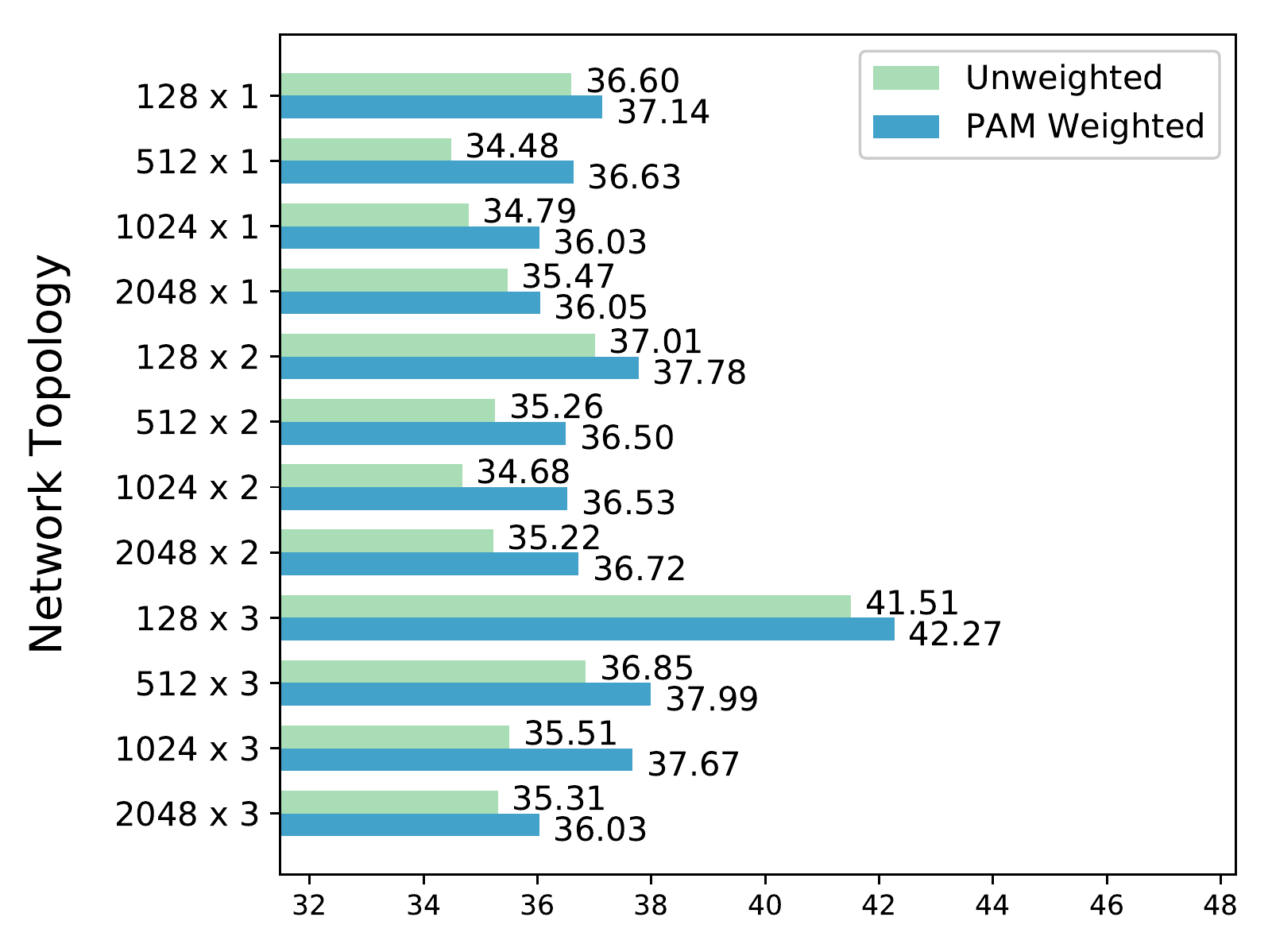}}
    \subfigure[IPS]{\includegraphics[height=1.8in,trim={0 0 .7cm 0}]{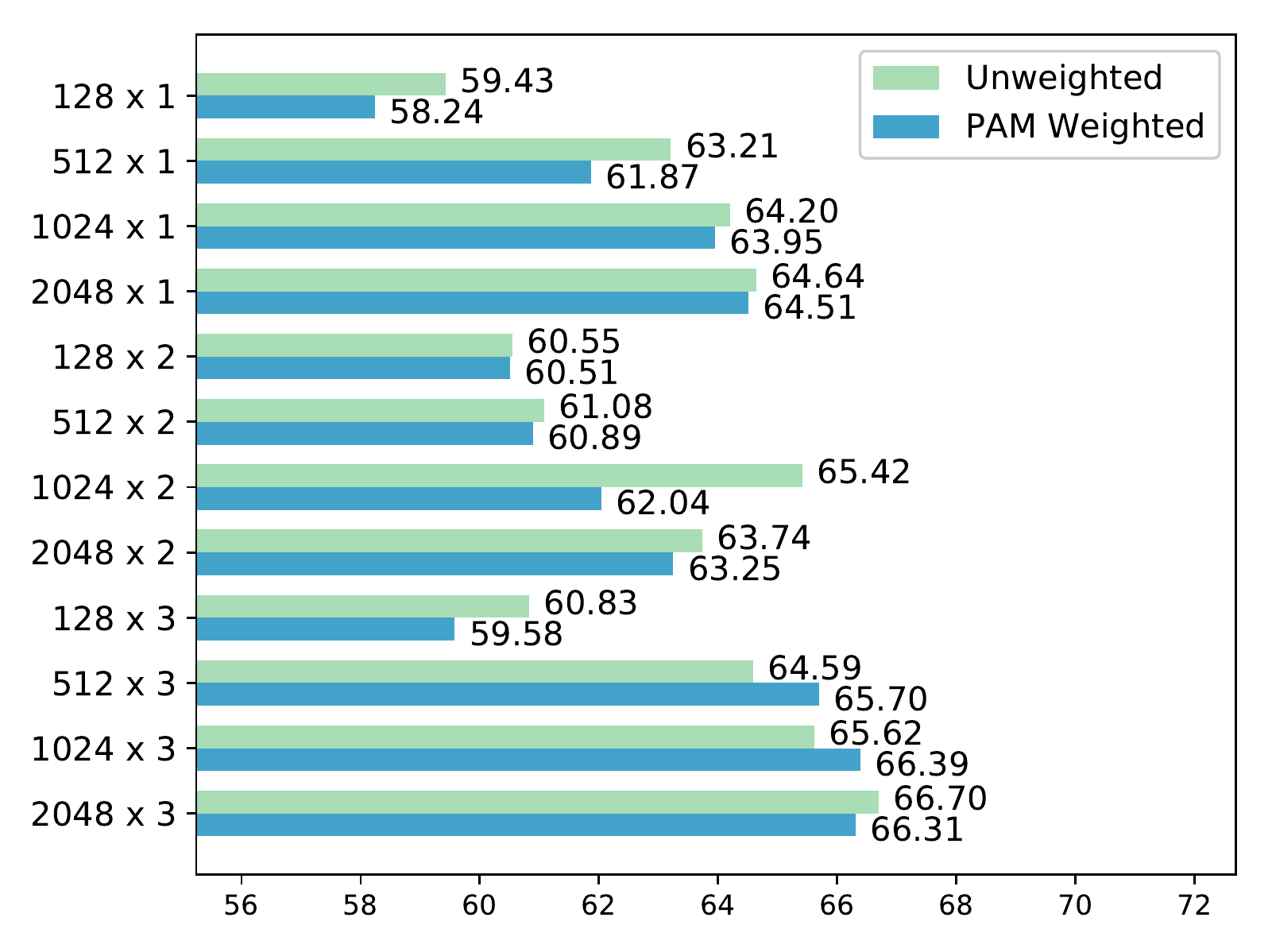}}
    \subfigure[APS]{\includegraphics[height=1.8in,trim={0 0 .7cm 0}]{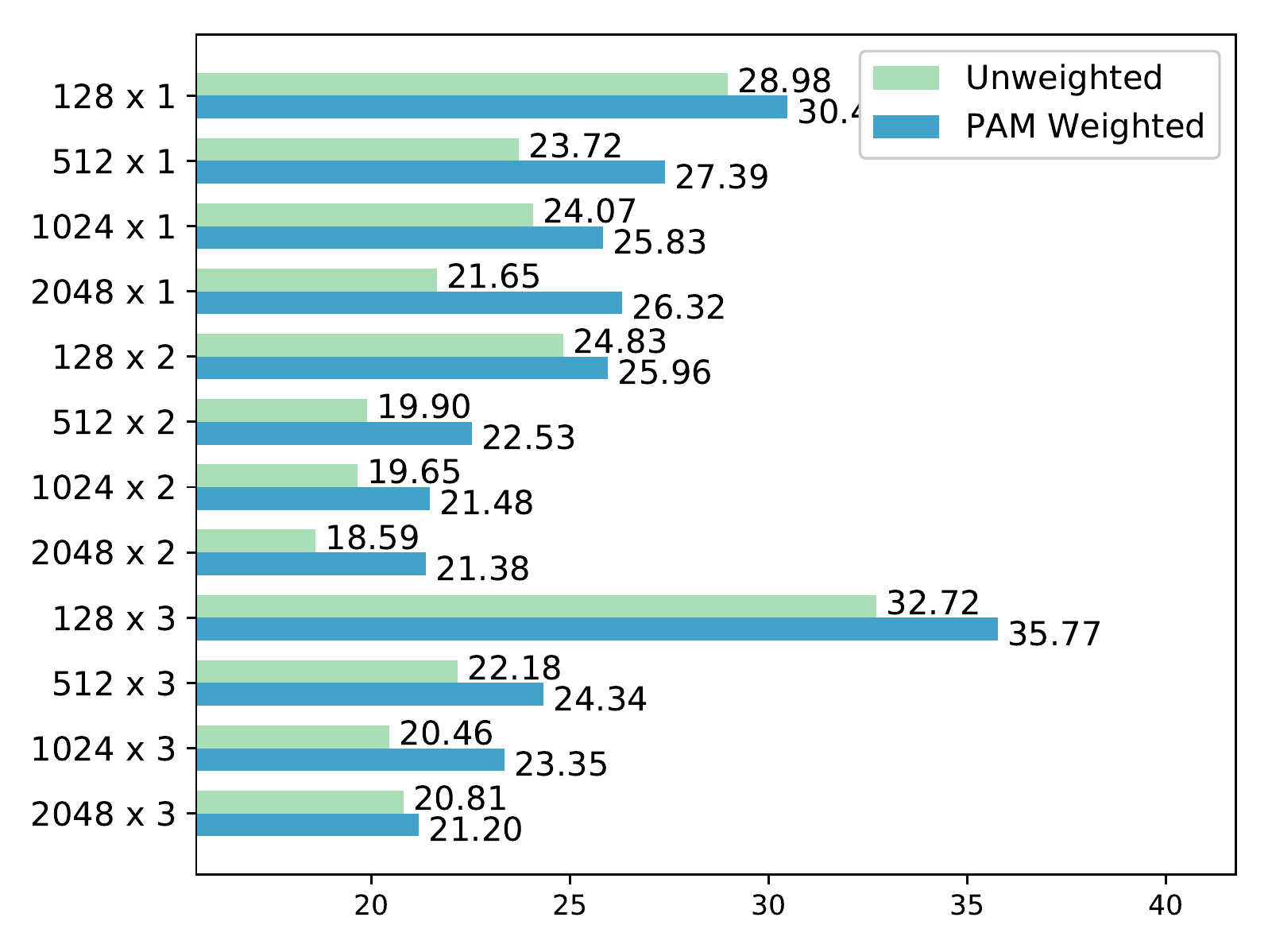}}}
    \caption{Comparison of BSS-Eval and PEASS Software scores.}
    \label{bars}
\end{figure*}

\section{Experimental Results}
\label{sec:results}

For the present, we use the BSS\_Eval toolbox to evaluate the objective sound quality \cite{VincentE2006ieeeaslp}, as well as the STOI and PEASS toolkits for the perceptual evaluation \cite{peass}\cite{taal2010short}. These separation metrics are used to compare the denoised speech results obtained from the proposed neural network, using either the conventional MSE cost function \eqref{loss1} or the perceptually weighted MSE cost function \eqref{eq:error-pam}.

\subsection{Objective Quality Assessment}
\label{bssres}


We use BSS\_Eval toolbox \cite{VincentE2006ieeeaslp} to objectively compare the denoising quality over two groups of models. Particularly, we consider three measures: SIR for the ratio of the source over the remaining interference, SAR to measure the amount of artifacts introduced during the separation process, and SDR to reflect the overall source separation performance. These measures are calculated for each reconstructed utterance and are presented in Fig. \ref{bars} as weighted averages over $400$ speech signals based on the lengths of the signals.
The weighted models overlook noise below the global masking threshold, focusing instead on audible noise affecting human speech perception. Because of this, the model does not objectively denoise the utterance, resulting in a slightly lower SIR (Fig.~\ref{bars} (b)) than the unweighted model. However, the perceptual models add barely any artifacts in comparison to the unweighted models (Fig.~\ref{bars} (c)). Similarly, the output SDR of networks utilizing the PAM weights is still comparable to that from conventional denoising networks (Fig.~\ref{bars} (a)). Irrespective of the cost function being weighted or unweighted, there is a trend--wider and deeper neural networks guarantee better objective separation quality. This trend justifies the popularity of deep learning for speech denoising tasks. However, we argue that the variation of the perceptual quality with respect to the model complexity has a different pattern.
 



\begin{figure}[t]
\centering
    \includegraphics[height=1.6in, trim={0 .7cm 0 .7cm}]{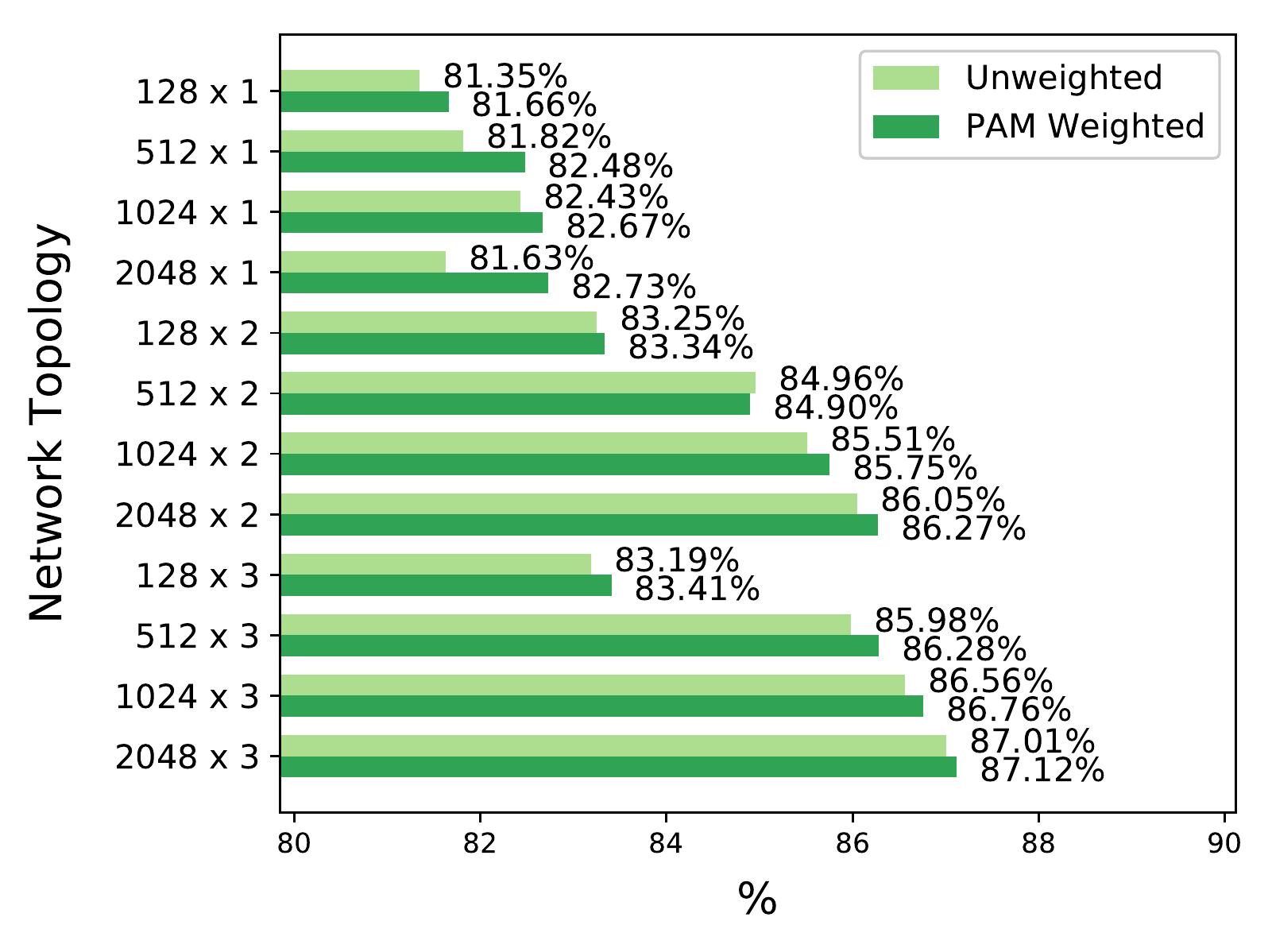}
    \caption{Comparison of Short-Time Objective Intelligibility scores.}
    \label{stoi}
\end{figure}

\subsection{Perceptual Quality Assessment}
PEASS measures the the perceptual quality of the denoised speech signals. In the case of single-channel source separation there are three metrics: Overall, Interference-related, and Artifact-related Perceptual Scores (OPS, IPS, and APS). These perceptual scores complement objective measures SDR, SIR and SAR, respectively. For all $12$ network topologies considered, the proposed perceptual weighting yields higher OPS (Fig. ~\ref{bars} (d)). 
While SDR is highly contingent on network complexity, OPS is well-maintained even by shallow networks. As discussed in Section ~\ref{bssres}, PAM-weighted networks do not equally consider all frequency bins. Therefore, the IPS is reasonably lower than that from non-PAM models (Fig. ~\ref{bars} (e)). However, artifacts introduced from perceptual weighting are also dampened which leads to higher APS (Fig. ~\ref{bars} (f)). Overall, we can conclude that the PEASS evaluation metrics says that the proposed perceptual weighting leads to compressed network structures that do not suffer from as much perceptual performance drop as the ones that minimize an unweighted MSE cost function. 

We additionally verify the effect of the proposed weighted cost function on Short Time Objective Intelligibility (STOI) scores \cite{taal2010short}. In Fig. ~\ref{stoi}, we see that the average STOI score from weighted models is marginally higher, which reassures the stability of the perceptual weighting scheme. However, we stay conservative in asserting that the proposed method improves speech intelligibility, a claim which can be researched further by performing subjective evaluation of audio quality with frameworks such as MUSHRA \cite{mushra}.

\section{Conclusion}
In this paper, we proposed a psychoacoustically weighted cost function that leads to a more efficient network structure for speech denoising tasks. Such efficient networks are with a less number of parameters, so that their implementations are hardware-friendly especially in the resource-constrained environments, while they maintain comparable perceptual quality. In the future, we plan to investigate the effect of the proposed perceptual weighting scheme on the other kinds of network compression such as the quantization scheme for the parameters.

\bibliographystyle{IEEEbib}
\bibliography{mjkim}

\end{document}